\documentstyle[aps,multicol,prl,epsfig]{revtex}

\newcommand{\be}{\begin{eqnarray}}
\newcommand{\ee}{\end{eqnarray}}

\newcommand{\r}{{\bf r}}

\begin{document} 

\draft

\title{The universality class  of fluctuating pulled fronts}

\author{Goutam Tripathy$^1$, Andrea Rocco$^{2,3}$, Jaume Casademunt$^3$, 
and Wim van Saarloos$^1$}
\address{
$^1$ Instituut--Lorentz, Universiteit Leiden, Postbus 9506,
  2300 RA Leiden, The Netherlands\\
$^2$ Dipartimento di Fisica, Universit\`a di Roma ``La Sapienza'', 
P.le Aldo Moro 2, I-00185 Roma, Italy\\
Istituto Nazionale Fisica della Materia, Unit\`a di Roma\\
$^3$ Departament ECM, Universitat de Barcelona, 
Av. Diagonal 647, E-08028, Barcelona, Spain\\
}

\date{\today} 

\maketitle

\begin{abstract}    
It has recently been proposed that
fluctuating ``pulled'' fronts propagating into an unstable state should 
not be in the standard KPZ universality class for rough interface
growth. We introduce an effective field equation for this class of
problems, and show on the basis of it  that noisy pulled fronts in
{\em d+1} bulk 
dimensions should be in the universality class of the {\em (d+1)+1}D
KPZ equation rather than of the {\em d+1}D KPZ equation. Our scenario
ties together a number of heretofore unexplained observations in the
literature, and is supported by previous numerical results.

\end{abstract} 


\begin{multicols}{2}

Consider spatio-temporal systems in which the important dynamics is
governed by the propagation of fronts or interfacial zones separating
two domains whose bulk dynamics is relatively trivial or
uninteresting. In the presence of fluctuations, the theory of the
stochastic behavior of such fronts or interfaces is well-developed \cite{review1,review2}. In
particular it is known that many such fluctuating $d$-dimensional
interfaces in {\em d+1} bulk dimensions are described by the KPZ
equation \cite{KPZ} for their height $h$,
\begin{equation}
\frac{\partial h}{\partial t} =\nu \nabla^2 h + 
\lambda (\nabla h)^2 + \eta ~, \label{kpz}
\end{equation}
with $\eta$ a random gaussian noise with correlations
\be
&&\langle \eta(\r_{\perp},t) \rangle = 0 \label{mean}~,\\
&&\langle \eta(\r_\perp,t) \eta (\r_{\perp}^{\prime},t^{\prime}) \rangle = 2
\epsilon \;
\delta^d (\r_{\perp}-\r_{\perp}^{\prime}) \delta(t-t^{\prime})~. \label{cor}
\ee
We will follow common practice to refer to this equation as the {\em d+1}D(imensional) KPZ equation, where the $d$ refers to the dimension of the
interface and the {\em +1} to the  time dimension; $\r_{\perp}$
denotes  the coordinates perpendicular to the direction of propagation
of the interface.

The fact that the scaling behavior of so many stochastic interfaces
fall in the {\em d+1}D KPZ universality class, is due to the fact that
{\em (1)} this equation contains all the terms in a gradient expansion
which are relevant in a
RG sense; and {\em (2)} that the long wavelength deterministic
dynamics of many interfaces is {\em local in space and time}, i.e., of the
form $v_n=v_n(\nabla h, \nabla^2 h,\cdots)$, expressing that the
normal velocity $v_n$ becomes essentially a function of the instantaneous slope
(angle)  and  
curvature of the interface only. Upon
expanding in the gradients, adding noise,  and retaining only relevant
terms, one then arrives at (\ref{kpz}).

The starting point of such an argument, the fact that one can
integrate out the internal structure of the interface and on long
length and time scales think of it as a mathematically sharp boundary
with effective dynamics expressed by a boundary condition 
$v_n$=$v_n(\nabla h, \nabla^2 h,\cdots)$ which is local in space and
time, is appealing and usually 
correct. Intuitively, one associates it with the interfacial zone
being sufficiently sharp on a spatial scale. Nevertheless, there have
been scattered observations in the literature which indicate that
there is more to it: {\em (a)} Some continuum reaction-diffusion
equations have propagating planar  interfaces of finite width which
are stable, but which become weakly unstable for discrete particle
model equivalents \cite{kl}, contrary to what the above coarse-graining
picture would suggest. {\em (b)} The empirical relation observed for the
distribution of DLA fingers in a channel and the interface shape of a
viscous finger could not be understood from the standard continuum
model until the innocuously looking reaction term was regularized \cite{blt};
on hindsight, this was because the standard mean-field DLA equations
do not give the appropriate ``local'' boundary conditions of the type $v_n=-\mu
\nabla_n p$.
 {\em (c)} In a simple 
stochastic  particle model with fluctuating fronts, non-KPZ scaling
was observed \cite{Riordan95} contrary to what one would naively have
expected. 

It turns out that these observations all have one common denominator
\cite{hl,Tripathy00}, in that they are related to the existence of two
classes of fronts, ``pushed'' and ``pulled'' fronts. {\em Pushed} fronts are
the usual ones: their dynamics is determined by the behavior in
the interfacial zone, a region of finite thickness, and their response
to the bulk fields is local in space and time
\cite{Ute00-1,Ute00-2}. {\em Pulled} fronts, on the other hand, propagate
into a linearly {\em unstable} state. Although they do not differ
noticeably from pushed fronts in
their appearance, their dynamics is driven by the
growth and spreading of perturbations about the unstable state in the
semi-infinite region {\em ahead} of the front \cite{Ute00-1}; hence
they are particularly
sensitive to slight changes in the dynamics there
\cite{bd,kl}. These important differences led two of us
\cite{Tripathy00} to propose
recently that fluctuating variants of $d$-dimensional pulled fronts in
{\em d+1} bulk dimensions would indeed {\em not} be in the {\em d+1}D KPZ
universality class, even though pushed fronts do effectively give
local boundary conditions on long length and time scales, and hence
{\em  do} 
give rise to {\em d+1}D KPZ scaling in the absence of coupling to a
diffusion or laplace field in the bulk  \cite{notedla}. Simulations
of a simple stochastic lattice model were consistent with  these
arguments, and with the earlier observations of \cite{Riordan95}. 

In this paper, we will argue that fluctuating pulled fronts are indeed
in a different universality class  from the usual pushed ones which
show the standard KPZ behavior. Indeed, we will show that the
semi-infinite region ahead of the front can {\em not} be integrated
out, and effectively enhances the dimension by {\em 1}: we introduce a field
equation for fluctuating pulled fronts and argue that
 $d$-dimensional fronts in {\em d+1} bulk dimension
are in the universality class of  the {\em (d+1)+1}D KPZ rather than
the {\em d+1}D KPZ equation. This surprising scenario, which also builds on
the insight of \cite{Rocco00-2} for the stochastic behavior of pulled
fronts in one bulk 
dimension,  is fully consistent with our
earlier {\em 2}D simulations \cite{Tripathy00} and also with
the heretofore unexplained results of \cite{Riordan95} in higher
dimensions. In addition, as we 
shall discuss, our scenario leads to a number of 
interesting new questions and challenges.

A stochastic equation for pulled fronts should obey two
requirements: in the usual stochastic lattice models with fronts, no
particles are spontaneously generated when there are none already.
Secondly, the average front speed and the local fluctuations ahead of the front remain always finite
in such lattice models \cite{note3}. The field equation should
be consistent with these basic facts. So when we consider a stochastic
field equation for 
 $\phi(x,\r_{\perp},t$ in {\em d+1} dimensions $(x,\r_{\perp})$ of the type
\begin{equation}
\label{FK}
\frac{\partial \phi}{\partial t} = D \nabla^2 \phi + f(\phi)
+ g(\phi)\eta \label{maineq}
\end{equation}
these requirements put constraints on the function $f$  and the
 noise term $g(\phi)\eta$. The stochastic
noise  $\eta(x,\r_{\perp},t)$  has  delta 
correlations as in (\ref{cor}), and is interpreted in the Stratonovich
sense, but our arguments will not rely on  the distinction between Ito 
and Stratonovich calculus.  Stochastic
field equations of this type have e.g. been used already since long
for studying 
the scaling behavior of {\em homogeneous} bulk  phases like directed
percolation \cite{dp}; investigations of noisy fronts in such
equations are more recent --- see e.g. \cite{Rocco00-1} for an
analysis of stochastic pulled fronts and a discussion
of the applicability to various systems. Here we focus on the proper form for an effective
stochastic field equation for pulled fronts. For $f(\phi)$,
which determines
the dynamics of deterministic fronts in the absence of noise, we
choose the standard form for pulled front propagation
$f(\phi)=\phi-\phi^3$, which gives saturation of the field $\phi$
behind the front where $\phi \to 1$. How should the noise term
$g(\phi)\eta$  look like \cite{notenoise}? The requirement that
if there are no particles ($\phi$=$0$) none are created spontaneously, 
implies that there should be no additive noise term,  and hence that
$g(\phi$=$0)=0$. For $\phi$ nonzero but small, it is natural to assume a
power law behavior $g(\phi)\sim \phi^\alpha$; in the studies of the
homogeneous bulk properties of 
directed percolation the choice $\alpha=1/2$ was been made \cite{dp}, motivated
by the idea that typical bulk fluctuations are of the order of the square
root of the particle density.  For pulled fronts, however, the
dynamically important region is ahead of the front, where $\phi
\to0$. Our second requirement that the relative fluctuations
$g(\phi)\eta /\phi$ remain finite here shows that the natural choice is  $\alpha=1$,
i.e., $g(\phi) \sim \phi $ for $\phi \ll 1$ \cite{note2}. The
linearity of $g$ 
for small $\phi$ is sufficient for our subsequent analysis. In
our numerical studies, we have actually taken
$g(\phi)=\phi(1-\phi^2)$, a form taken to suppress fluctuations behind
the front. This makes it numerically easier to focus on the
fluctuations of the front position itself, without affecting the essential
results.  Specifically, we thus propose as the generic stochastic
field equation for pulled fronts
\begin{equation}
\label{FK2}
\frac{\partial \phi}{\partial t} = D \nabla^2 \phi + (1+\eta ) \phi
(1-\phi^2)~.\label{maineq2}
\end{equation}

Let us now turn to the analysis of the stochastic behavior of 
fronts which propagate along the $x$-direction into the linearly
unstable state $\phi$=$0$.  The crucial feature of {\em pulled} fronts is
that even though the full dynamics of the fronts is nonlinear, it is
essentially determined in the ``leading edge'', the region ahead of
the front where $\phi$ remains small enough that the nonlinear
saturation term $-\phi^3$ which limits the growth, plays no role: the
linear spreading and growth of perturbations about the state $\phi$=$0$ almost
literally ``pull the front along''. An important recent development
has been the realization that   this simple
intuitive picture can be turned into a systematic scheme to calculate
even the convergence of the front speed to its asymptotic value
$v^*$. Remarkably, this relaxation is governed by universal power laws
which can be calculated exactly even for general equations
\cite{Ute00-1}. The fact that the 
stochastic fluctuation effects that we want to investigate are
dominant relative to the deterministic velocity relaxation terms,
suggests to calculate these along similar lines.
For the deterministic case ($\eta$=$0$), fronts in 
(\ref{FK2}) propagate with an asymptotic speed $v^*=2\sqrt{D}$. In a
frame $\xi=x-v^*t$ moving with this  speed, the asymptotic front
solution has an exponential fall-off  $\sim e^{-\lambda^* \xi}$ with
$\lambda^*= 1/\sqrt{D}$ for large positive $\xi$. The asymptotic
relaxation analysis of 
deterministic fronts is based on the so-called leading edge transformation
$\phi=e^{-\lambda^*\xi} \psi$, which transforms (\ref{FK2}) into
\begin{equation}
\label{FK3}
\frac{\partial \psi  }{\partial t} = D \nabla^2 \psi  + \eta \psi
 - (1+\eta)   \psi^3 e^{-2\lambda^*\xi}~.\label{maineq3}
\end{equation}
Here $\psi$ has been written in the frame moving with velocity $v^*$
in the $x$-direction: $\psi=\psi(\xi,\r_{\perp},t)$.

In the analysis of deterministic fronts \cite{Ute00-1}, the nonlinear
term on the right hand side (which is exponentially small for
$\xi \gg 1$) essentially plays the role of a boundary condition  for
the semi-infinite leading edge region where $\phi$ is small --- it
allows the nonlinear region to match properly to the leading edge
which ``pulls'' the front.  As explained above, this holds {\em a
fortiori} for fluctuating pulled fronts: their stochastic
fluctuations are essentially determined by the region where the
linearized equation can be used. Now, as is well known, upon
making a Cole-Hopf transformation $\psi=e^h$, the linearized equation
transforms to
\begin{equation}
\frac{\partial h}{\partial t} =D \nabla^2 h + 
D (\nabla h)^2 + \eta ~, \label{kpz2}
\end{equation}
which is nothing but the {\em (d+1)+1}D KPZ equation (\ref{kpz}) for
the {\em d+1} dimensional field $h(\xi,\r_{\perp},t)$!

\begin{figure}
\begin{center}
\epsfig{figure=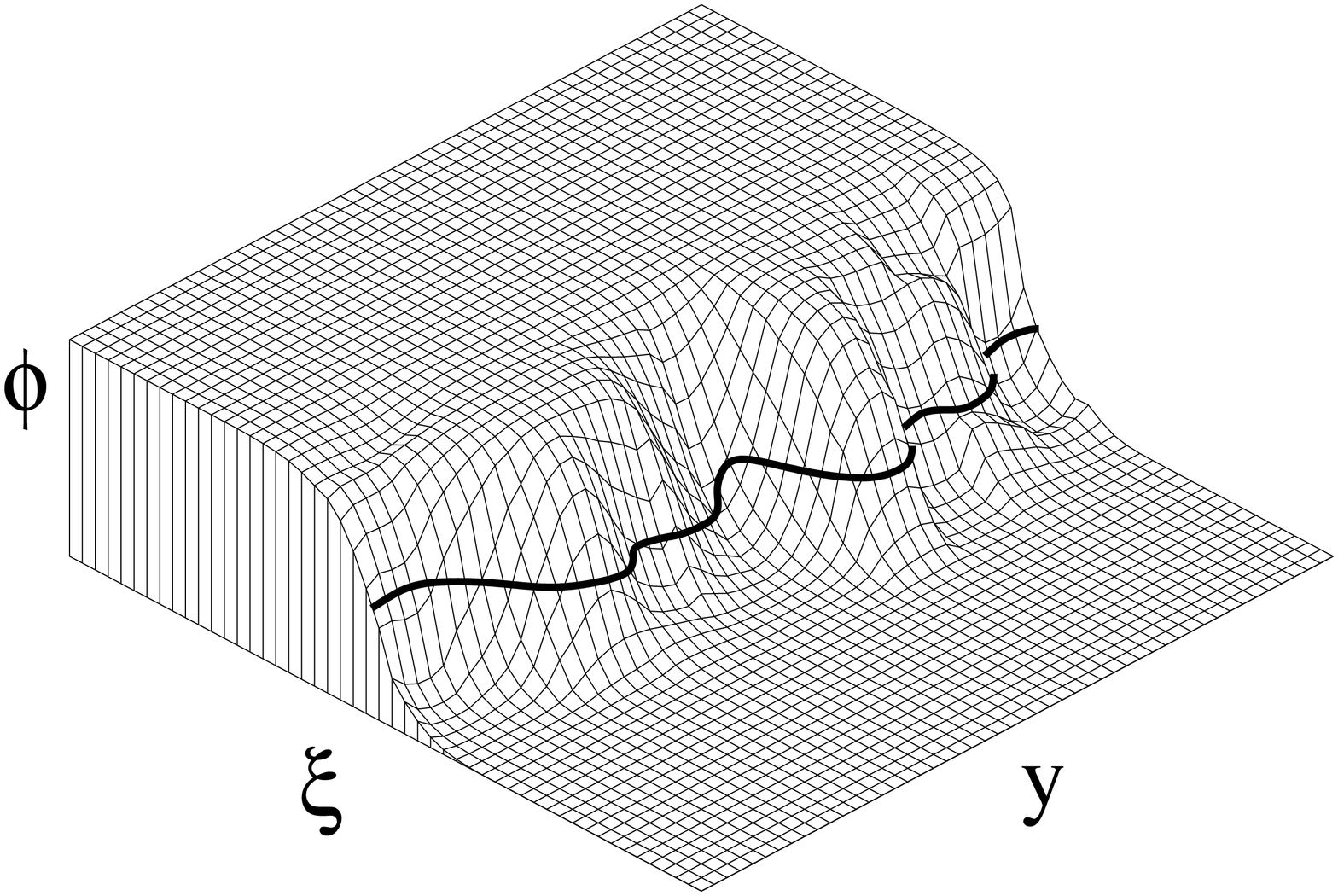,angle=-90,width=0.465\linewidth}\hspace*{2mm}
\epsfig{figure=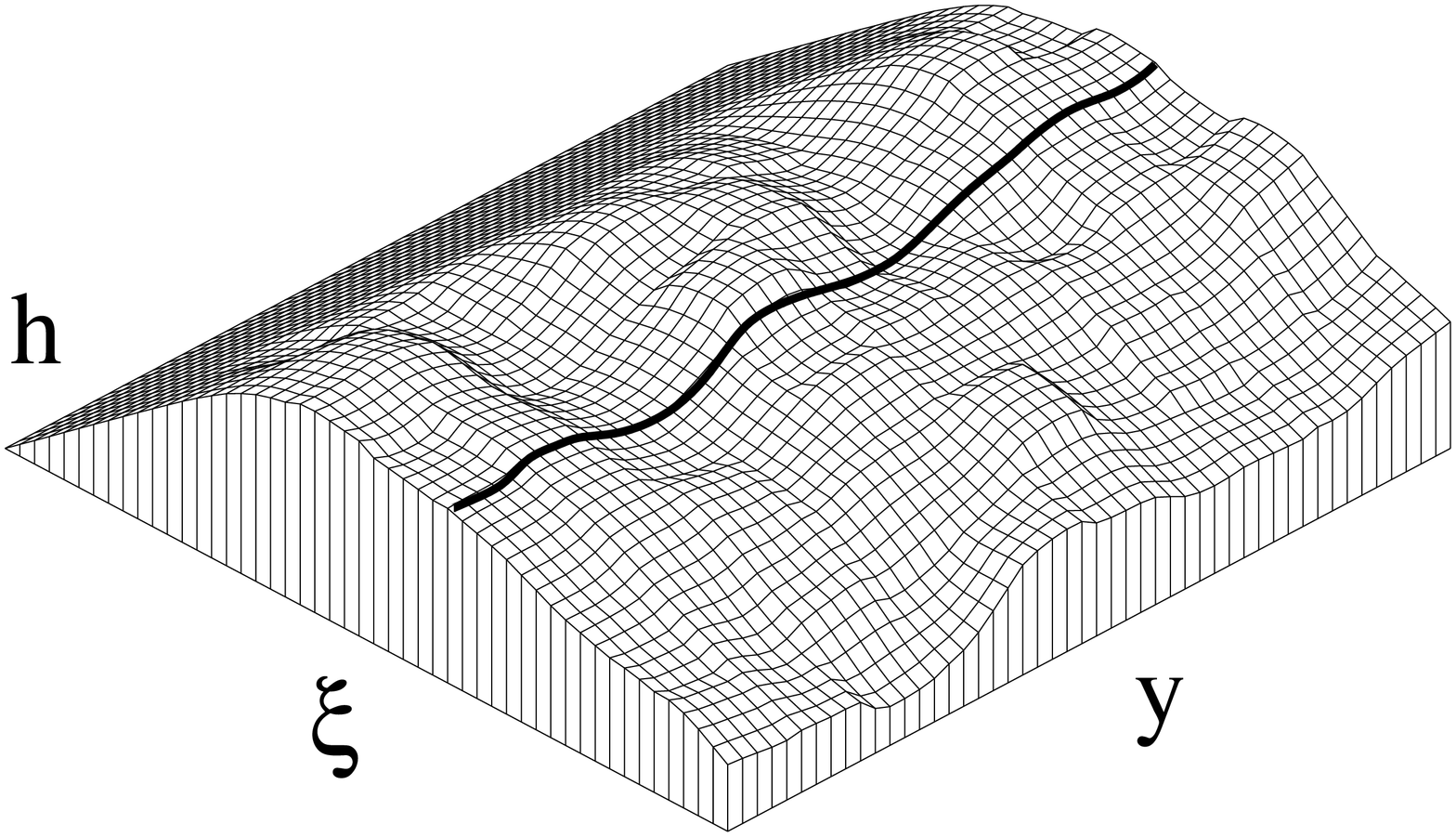,angle=-90,width=0.465\linewidth}
\end{center}
\caption{Left panel: snapshot of the field  $\phi$ at time $t=20$ in a
2D simulation of 
(\ref{FK2}) with $D=1$ and $\epsilon=10$. The thick line is the position of the
front, defined by tracking the line where $\phi(\xi,r_{\perp},t)=1/2$. Right
panel: the same data as in the left figure, plotted in terms of the
height variable  $h$.  Note that $h$ has the appearance of a (slanted)
fluctuating surface. The flat portion on the left is the region behind
the front  and where $h \approx \lambda^* \xi$ since $\phi \to 1$. The
thick line 
indicates the height fluctuations 
along a line of constant $\xi$. This illustrates that the
one-dimensional {\em position fluctuations} along the pulled front
illustrated by 
the thick line in the left panel are related to the {\em height 
fluctuations} of the two-dimensional fluctuating surface of the
leading edge variable $h$. The scaling behavior of these is that of
the  {\em 2+1}D KPZ universality class. 
} \label{phihplot}
\end{figure}

As illustrated for two bulk dimensions in Fig. \ref{phihplot}, the 1D
fluctuations in the front position in the propagation direction are
defined by tracing a line where $\phi= const$, e.g., $\phi=1/2$. Since
$\phi=e^{-\lambda^* \xi +h}$, the front fluctuations in the $\xi$
direction are given by  $\xi(\r_{\perp},t) = h(\xi, \r_{\perp},
t)/\lambda^* +\xi_0$ $\approx$ $  h(\xi_0, \r_{\perp},
t)/\lambda^* +\xi_0$, where the constant $\xi_0$ is determined by the
level curve of $\phi$ which we trace to determine the front position.
 Thus indeed the position fluctuations of a $d$-dimensional pulled
front in {\em d+1} bulk dimensions map onto the height fluctuations
along a line of a
KPZ surface in {\em d+1} dimensions --- see Fig. \ref{phihplot} The
growth and roughness exponents  
are therefore  those of  the {\em (d+1)+1}D KPZ universality class!

The above scenario unifies a number of different results. It can
immediately be compared with the simulation 
results of the  stochastic lattice model of
\cite{Tripathy00}. In that paper a {\em 2}D lattice model was introduced in
which by changing a simple birth and death rule of particles {\em 1}D fronts
could be tuned from pushed to pulled. The scaling exponents of the
pushed model were found to be the standard {\em 1+1}D KPZ ones, as it
should, while those of the pulled variants were close to those of the
{\em 2+1}D KPZ universality class. More importantly, without any
adjustable parameters, the distribution functions for the long-time
saturated width  of the fronts in this model for finite transverse
width $L_{\perp}$ \cite{racz} are completely in accord with our scenario \cite{Tripathy00}.
Moreover, although fronts in {\em 1}D do not have transverse
fluctuations, the wandering of the position of pulled fronts in one
dimenension is
also consistent with {\em 1+1}D KPZ scaling \cite{Rocco00-2}. Finally, the observations
of Riordan {\em et al.} \cite{Riordan95} that in three (and higher)
bulk dimensions 
their fronts did 
not appear to show a power law growth of the front  width finds a
natural explanation. According to \cite{Tripathy00} their fronts are
pulled and so  they should be governed by
the {\em 3+1}D KPZ equation. The free $\lambda=0$ fixed point in this
equation is stable and has no divergent interface width. Apparently 
above two dimensions  the model of \cite{Riordan95,Tripathy00}
is in the weak-coupling limit.

On hindsight, our arguments also justify the regularization of
\cite{blt} of the mean-field equations for DLA in a channel: the full
problem involves  pushed fronts but the
mean-field equations have pulled front solutions. The regularization
effectively cures this by making the fronts into pushed ones.


\begin{figure}[h]
\begin{center}
\vspace*{-0.4cm}

\epsfig{figure=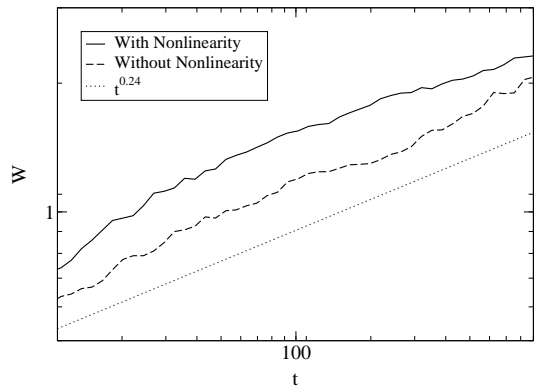,angle=-90,width=0.8\linewidth}
\end{center}
\caption[]{The increase of the root mean square front width  
$W = [ \langle\overline{[h(\r_{\perp},t)-\overline{h}(\r_{\perp},t)]^2}\rangle]^{1/2 }$ 
(with the overbar denoting an  average over ${\bf r}_{\perp}$)
as  a function of time. Data are for
simulations of Eq. (\ref{FK3})  both with nonlinearity (full line) and
without nonlinearity (dashed line), for $\epsilon=5$ and an
 effective diffusion
constant $D=0.4$, which
corresponds to dimensionless KPZ coupling constant $\tilde{\lambda}=
25$ \cite{notedeff}. The front position $h$ in the $x$ direction is defined as the level line
where  $\phi=0.5$. The fact that the growth exponent is
essentially the same with and
without nonlinearity in the $\phi$ equation justifies our assertion
that these terms do not affect the dominant scaling behavior of pulled
fronts.}\label{bothplots}
\end{figure}

The validity of the crucial step of our derivation, the assertion that
the nonlinearities in (\ref{FK2}) or (\ref{FK3}) can be neglected
because the leading edge where $\phi$$\ll$$ 1$ is the essential region,
can be tested independently. In Fig. \ref{bothplots} we show
simulation data of the wandering of the lines where $\phi= 0.5$ in Eq. (\ref{FK2}) in {\em 2}D, both with and without the
nonlinearity. Following \cite{Beccaria94}, where
 the linearized version of (\ref{FK3}) was already employed to study the
{\em 2+1}D KPZ exponents numerically, we have taken parameters so as
to make the dimensionless coupling  $\tilde{\lambda}= 2 \lambda^2
\epsilon / \nu^3 \approx 25$. This value appears to be close to the fixed
point value and so slow transients are minimized \cite{Beccaria94}. We
find indeed that the two datasets 
with and without the nonlinear term in (\ref{FK2}) show the {\em same}
growth exponent, with a value close to the one $\beta \approx 0.24$ of the
{\em 2+1}D KPZ equation. This gives confidence in the validity of our assertion
that the nonlinear terms in the front equation are not important for
the scaling behavior of pulled fronts.

The main steps of our line of argument are elegantly
direct and build on various previously established ideas; at the same
time our
scenario also raises a number of new questions and challenges for
further research:\\
{\em (i)} There is no systematic theory for the transition from the
pushed to the pulled regime in stochastic lattice models, so it is
difficult to determine {\em a priori} which models lead to the
standard  pushed fronts and
which ones to pulled ones. E.g., fronts in the directed percolation
problem are pushed  and obey KPZ scaling in one special case
\cite{dhar}, but it 
is not known whether this is generally so.\\
{\em (ii)} Finite size scaling of the KPZ equation is normally done for
interfaces of size $L_{\perp}$ in all directions. Our scenario, on the other
hand, leads one to consider  anisotropic scaling, since there is
effectively a time-dependent cutoff  in the $\xi$-direction
\cite{Rocco00-2}.  
The crossover scaling is completely
unexplored, but is most likely quite tricky: for fixed
$L_{\perp}$ the results of \cite{Rocco00-2} for fronts in one
dimension suggest that one should see subdiffusive wandering of the
average front position, $\langle (\overline{h})^2 \rangle
\sim \sqrt{t}$ (rather than $\sim t$) because the cutoff in the
$\xi$-direction grows as 
$L_\xi \sim \sqrt{t}$, but our simulations seem to suggest that the
crossover to this regime happens at such extremely long times that it
can not convincingly be seen in practice. Moreover,  the crossover is
likely to depend significantly on the initial conditions
\cite{Rocco00-2}. \\
{\em (iii)} According to the results of \cite{bd,kl,hl}, pulled fronts
are very sensitive to finite particle effects, so that the convergence
to a continuum limit is extremely slow. This crossover is  poorly
understood, especially in higher dimensions where it is not even known
how important the effects are in practice.

In conclusion, we have put forward an effective field equation for
pulled fronts  and argued on the basis of it that pulled fronts in
{\em d+1} bulk dimensions are in  the {\em (d+1)+1}D KPZ
universality class rather than the {\em d+1}D KPZ universality
class. The scenario ties together various results in the literature
and brings up various new issues for future research.

WvS would like to thank Uwe Tauber and David Mukamel for stimulating
discussions. 
Financial support from the Dutch science foundation FOM, the Spanish
project  BXX2000-0638-C02-02 and the  TMR network ERBFMRX-CT96-0085 is
gratefully acknowleged.

\end{multicols}


\begin{references}
\bibitem{notedla} The Laplacian growth model \cite{review2} or DLA in
a channel \cite{blt} are
examples of models with local boundary
conditions where the coupling to the Laplacian
diffusion field gives non-KPZ behavior.
\bibitem{note3} The requirement that  
 the  fluctuations in $\phi$ are of order
one in the leading edge is in line with the ideas of Brunet and Derrida \cite{bd} for the
effect of a finite particle cutoff.
\bibitem{notenoise}
Actually, as is well known, the Stratonovich 
field equation is only well defined if the 
 noise correlation length is made finite by writing  $\langle
\eta(\r,t) \eta (\r^{\prime},t^{\prime}) \rangle = 2 \epsilon/
\Lambda^d C(|\r-\r^{\prime}|/\Lambda) \delta(t-t^{\prime}) $, with
$\int d\r \; C(\r)=1$. This cutoff does not affect the universal
long time scaling.
\bibitem{note2} Although our argument is equally valid in the Ito
case, we note that  in the Stratonovich case, the finiteness of the
relative fluctuations and of the renormalized front speed are
connected. Indeed,  the
fluctuation average of the multiplicative noise term is then $\langle
g(\phi)\eta\rangle = \epsilon \Lambda^{-d} C(0)  \langle g' (\phi) g(\phi)\rangle$
(see e.g. \cite{Rocco00-1}). Here 
$C(0)$ is the value in the origin of the regularization of the
delta-function, see \cite{notenoise} above. For $g \sim \phi^\alpha$,
this gives a term proportional to $\phi^{2\alpha -1}$, which for
$\alpha<1$ leads to an infinite front velocity. Note also that for
$\alpha>1$ the  analysis in the text implies that the pulled front is
unaffected  by the noise, so only the choice $\alpha=1$
is consistent.
\bibitem{notedeff} Our simulations were done on a grid with spacing
$\Delta x=1$ and bare diffusion constant $D_0=0.1$. According to
section 5.6.6 of \cite{Ute00-1} this translates into an effective diffusion
coefficient $D=0.40039$ in the $\psi$-equation (\ref{FK3}). Our
spatial noise
correlation function $C$ of (see footnote \cite{notenoise}) is a
linear function 
that goes to zero at $\Lambda=5 $ gridpoints.
\end{references}
\end{document}